# Android Malware Detection using Machine learning: A Review

Md Naseef-Ur-Rahman Chowdhury[1], Ahshanul Haque[1], Hamdy Soliman[1], Mohammad Sahinur Hossen[1], Tanjim Fatima[1], and Imtiaz Ahmed[1]

New Mexico Tech, 801 Leroy PL, Socorro, NM, USA,
naseef.chowdhury@student.nmt.edu ahshanul.haque@student.nmt.edu,
hamdy.soliman@nmt.edu, mohammad.hossen@student.nmt.edu,
tanjim.fatima@student.nmt.edu, imtiaz.ahmed@student.nmt.edu

Abstract. Malware for Android is becoming increasingly dangerous to the safety of mobile devices and the data they hold. Although machine learning(ML) techniques have been shown to be effective at detecting malware for Android, a comprehensive analysis of the methods used is required. We review the current state of Android malware detection us ing machine learning in this paper. We begin by providing an overview of Android malware and the security issues it causes. Then, we look at the various supervised, unsupervised, and deep learning machine learning approaches that have been utilized for Android malware detection. Addi tionally, we present a comparison of the performance of various Android malware detection methods and talk about the performance evaluation metrics that are utilized to evaluate their efficacy. Finally, we draw atten tion to the drawbacks and difficulties of the methods that are currently in use and suggest possible future directions for research in this area. In addition to providing insights into the current state of Android malware detection using machine learning, our review provides a comprehensive overview of the subject.

Keywords: Android malware, mobile security, machine learning, detec tion, supervised learning, unsupervised learning, deep learning, perfor mance evaluation, comparison, limitations, challenges, future research directions

## 1 INTRODUCTION

Android malware attacks have skyrocketed in recent years due to the widespread use of mobile devices. Android malware is malicious software that targets security holes in Android devices. Malware for Android devices has the potential to harm one's financial situation as well as gain unauthorized access to personal information. As the number of Android malware attacks continues to rise, the importance of having reliable detection methods grows.

The well-established field of computer science known as machine learning has shown great promise for detecting Android malware. Because they can recognize complex data patterns and learn from large datasets, machine learning algorithms are ideal for detecting Android malware. Due to the growing interest in utilizing machine learning techniques for Android malware detection, numerous studies have been published in this area.

However, due to the scattered nature of the existing studies in this field, a comprehensive review of the machine learning-based ap proaches utilized for Android malware detection is required. This paper fills this void by providing a review of the current state of the art in Android malware detection using machine learning. In our re view, we will go over each of the various machine-learning techniques used to detect Android malware, the metrics used for performance evaluation, and the drawbacks and difficulties of the methods cur rently in use. We will identify future research directions for this field in the final section.

The purpose of this paper is to provide a comprehensive analysis of how Android malware is detected using machine learning. The approaches used, performance evaluation, potential drawbacks, and directions for future research will all receive special attention.

The operation of machine-literacy styles to the discovery of An droid malware is the sole focus of this disquisition. The study focuses on the following machine literacy-grounded aspects of Android mal ware discovery:

– An overview of Android malware and its security pitfalls. – Examination of the colorful supervised, unsupervised, and deep literacy machine literacy strategies employed for the discovery of malware on Android.

– Evaluation of the colorful machine learning styles used to descry malware on Android challenges and limitations of current styles, as well as openings for enhancement.

– Directions for unborn exploration in this area and suggestions for work to be done in the future.

Our exploration examines the current state of the art and the operation of machine literacy styles to the discovery of Android mal ware.

The remainder of the paper is structured as follows. Section 2 includes the existing literature review, section 3 depicts our method-ology, outcome and discussion introduced in section 4; then our con clusion is stated in section 5.

## 2 LITERATURE REVIEW

### 2.1 Overview of the Relevant Research

Due to the growing number of Android devices and the associated security risks posed by Android malware, the field of Android mal ware detection using machine learning has seen significant growth in recent years. For the purpose of detecting Android malware, super vised learning, unsupervised learning, and deep learning strategies have all been proposed by researchers[24].

Support vector machines (SVMs) and decision trees, two exam ples of supervised learning techniques, have been extensively utilized in Android malware detection[25]. In order to construct a model that is capable of distinguishing between legitimate and malicious Android applications, these methods rely on labeled training data.

Android malware detection has also utilized unsupervised learn ing techniques like clustering and dimensionality reduction. These techniques are capable of recognizing patterns in the data that may indicate malware and do not require labeled training data[26].

For Android malware detection, it has been demonstrated that deep learning techniques like Convolutional Neural Networks (CNNs) and Recurrent Neural Networks (RNNs) are effective[27]. When com pared to conventional ML approaches[27], these methods can boost malware detection accuracy by utilizing deep neural networks to ac quire intricate data representations.

Android malware detection has also utilized signature-based, rule based, and heuristic-based techniques in addition to ML methods[13][14]. However, the use of machine learning techniques for Android mal ware detection is the subject of this survey.

## 2.2 Classification of the existing Approaches

Various criteria, such as the type of learning, the features used, and the performance evaluation metrics used, can be used to classify the various machine-learning approaches used to detect Android mal ware.

There are two main types of machine learning approaches for Android malware detection, according to the type of learning: su pervised and unsupervised[26]. Unsupervised learning methods do not require labeled training data to construct a model, whereas su pervised learning methods do.

Machine learning methods for Android malware detection can be further categorized into the following groups according to the features they employ[28]:

Methodologies based on static analysis: These methods make use of features like the permissions that an Android application asks for and its code structure that are taken from static analysis.

Methods that are based on dynamic analysis: These meth ods make use of characteristics gleaned from the dynamic analysis of Android applications, such as the patterns of network communi cation and the application's behavior when it is running on a device.

Alternative methods: For Android malware detection, these strategies employ a mix of static and dynamic analysis-based fea tures.

There are several categories of machine learning approaches for Android malware detection based on the metrics used for perfor mance evaluation, including:

Methods based on accuracy: Precision, recall, and the F1-score are some of the accuracy metrics on which these methods base their evaluations of the machine learning model's performance.

Time-based methods: Time metrics, such as the amount of time needed to build the model and make predictions are used in these approaches to assess the machine learning model's performance.

Approaches based on robustness: The robustness of the ma chine learning model to adversarial examples, such as samples of malware designed to evade detection, is evaluated using these meth ods.

In summary, a clear understanding of the various machine-learning approaches used for this task and the criteria used to evaluate their performance is provided via the classification of the approaches used for Android malware detection based on the type of learning, the features used, and their performance evaluation metrics.

## Comparison of the Approaches

The authors in [1] present a new deep learning-based approach to detecting Android malware. The authors aimed at improving the accuracy and efficiency of Android malware detection by utilizing deep learning techniques. They utilized the Convolutional Neural Network (CNN) and Long Short-Term Memory (LSTM) ML algo rithms along with 20,00 android APK (10,000 benign and 10,000 malware). The results showed that the proposed system achieved high accuracy, with a value of 97.12%. The results also showed that the deep learning-based approach outperformed the traditional ML approaches in terms of accuracy.

The authors in [2] focused on the use of deep neural networks for the attribute-based recommendation. An input layer, hidden lay ers, and an output layer make up the multiple layers of the utilized deep neural network algorithm. The output layer predicts a score that indicates the likelihood that the user will prefer the item, after receiving user-item attributes from the input layer.

A real-world movie dataset containing information about users, movies, and their ratings was used in the experiments. The preci sion, recall, F1-score, and mean average precision are some of the evaluation metrics used to assess the proposed recommendation sys tem's performance. The effectiveness of using deep neural networks for attribute-based recommendation is demonstrated by the fact that the proposed algorithm outperforms other traditional recommenda tion algorithms in terms of precision and recall.

The authors in [3] suggested a strategy for engaging in adver sarial attacks on trading agents that are based on deep reinforce ment learning. The authors tested their method through its paces in two distinct trading settings: synthetic and historical datasets of the stock market. A reinforcement learning algorithm is used to teach a deep neural network to make trades based on market conditions. The authors then modify the decisions made by the agent by adding adversarial perturbations to the market state.

The results demonstrate that adversarial attacks can significantly affect the performance of deep reinforcement learning-based trad ing agents. The performance metric used is the profit or loss of the agent's trades. The adversarial attacks were successful in some in stances, but they were unsuccessful in others, yet resulting in profits.

In their conclusion, the authors state that reinforcement learning based trading agents must be robust.

The authors in [4] aimed at a comprehensive understanding of Android malware's characteristics and evolution. In order to iden tify common patterns and behaviors of malware, the authors investi gated a large dataset of Android malware and a benign applications' dataset. Additionally, they investigated the development of Android malware overtime to comprehend how it has advanced and changed. Though the authors claimed high accuracy results, the paper does not specify the quantitative metrics used for performance evaluation. Moreover, the paper does not clearly mention the utilized algorithms.

[5] is presented in "Virus Detection and Alert for Smartphones"[34]. The authors presented a system that is capable of detecting mal ware on a smartphone in real time and

letting the user know about it. However, though the authors claimed they have used dynamic analysis and mentioned high-accuracy results, the paper does not clearly mentions the utilized algorithm, exact accuracy results, and evaluation metrics.

The authors in [6], presented PUMA (Permission Usage to detect Malware in Android), a novel strategy for detecting malware on An droid devices. The authors contend that malware's excessive use of permissions can serve as a detection signature for malicious appli cations. The PUMA employs an ML-based algorithm that trains a classifier from a dataset (more than 4000 APKs containing both be nign and malware) of malware and benign apps. The app-requested permissions and their usage patterns are the features used for the classification. The authors stated that PUMA detects malware with an accuracy of over 90% and a low rate of false-positives.

In [7], a virus detection system based on data mining techniques is presented. The authors contend that large software datasets can be mined for patterns and features that can be used to identify malware.

The virus detection system's algorithm is not described in the pa per. However, the authors claim that they identify malware-inducing patterns and characteristics by employing data mining methods like the association rule of data mining and the ML decision trees algo rithms.

The paper does not specify the data used to evaluate the virus de tection system's performance. However, the authors claim that they evaluated a large dataset of software, which includes both beneficial and harmful software.

The paper does not specify the performance metric used to evalu ate the results. However, the authors assert that their virus detection system has a low rate of false-positives and high accuracy in identi fying malware.

The behavior of modern malware in the presence of anti-virtualization and anti-debugging techniques is the subject of the study in [8]. The authors argue that in light of the growing threat posed by malware, these methods, which are used to detect and prevent malicious ac tivity, have become increasingly important.

The behavior of malware in the presence of anti-virtualization(AV) and anti-debugging(AD) techniques is thoroughly examined by the authors. They evaluated the behavior of each sample when it is run ning in a virtual environment and when it is being debugged using a dataset of real-world malware samples. In addition, a classification framework is developed by the authors to classify the various AV and AD behaviors that were observed in the malware samples.

A dataset of actual malware samples was used in the study. The classification framework's ability to accurately classify the various kinds of AV and AD behaviors is the performance metric used to evaluate the results.

The study reveals a wide range of anti-virtualization and anti debugging behaviors in contemporary malware. The authors also find that these actions are getting better and more sophisticated, making it hard for anti-malware methods to stop them.

In [9], a singular value decomposition (SVD) method for detect ing metamorphic malware was presented. The authors evaluated the method's effectiveness with a large data set of benign and metamor phic executables.

The paper's algorithm is based on SVD, a mathematical method for looking at how data is structured. The singular values extracted from the executables' opcode sequences are used as features in an ML classifier, employing SVD. Control flow graph (CFG) and opcode n-gram analysis are two examples of traditional dynamic analysis methods that compare the efficacy of peers' works.

The experiments used a large collection of benign and metamor phic executables from a variety of sources as their data. The accu-racy, false-positive, and false-negative rates were some of the metrics used to evaluate the SVD-based method's performance. With an accuracy of 94.2% and a false-positive rate of 0.7 per cent, the SVD-based method performed better than conventional dynamic analysis methods[9]. The authors came to the conclusion that metamorphic malware can be effectively detected with SVD. In [10], a novel strategy for synthesizing malware specifications from suspicious behaviors is presented. The goal of the authors is to solve the problem of finding malware in large, complicated software systems, where traditional signature-based methods are frequently insufficient.

Through dynamic software system analysis, the authors deduced a novel algorithm for synthesizing malware specifications from suspi cious behaviors. A cost model and the findings of dynamic analysis are combined by the algorithm to produce near-optimal malware specifications in terms of coverage and specificity.

Software systems and their dynamic analysis results formed the data used in the study. They used the accuracy metric to evaluate their algorithm's performance. Such accuracy measure is also mea sured in terms of the synthesized malware specifications, measured in terms of both coverage (the proportion of malicious behavior that is detected) and specificity (the proportion of benign behavior that is not detected).

The study demonstrates that the proposed algorithm is capable of synthesizing malware specifications that are close to optimal for suspicious behavior. In addition, the algorithm outperforms conven tional signature-based methods in terms of accuracy[29], indicating its potential for enhancing malware detection in large, complex soft ware systems.

The authors in [11] presented a new approach for detecting mal ware on end-user devices. The authors propose a system that inte grates multiple techniques for detecting malware, including signature based detection, behavioral analysis, and data mining, to achieve improved accuracy and efficiency in comparison to traditional meth ods.

The authors use a combination of dynamic and static analysis techniques to extract features from malware specimens and build models that are used to detect malware on end-user devices. The performance of the system is evaluated using a large dataset of be-nign and malicious software, and the results show that the system is able to detect malware with high accuracy while incurring low overhead.

The algorithm used in the study is a combination of signature based detection, behavioral analysis, and data mining. The data used in the study consists of a large dataset of benign and malicious soft ware specimens. The performance metric used to evaluate the results is the accuracy of the malware detection system, measured in terms of the proportion of benign and malicious software specimens that are correctly classified.

The results of the study show that the proposed system is effective and efficient in detecting malware on end-user devices. The authors also find that the system outperforms traditional methods in terms of accuracy and efficiency, demonstrating its potential for improving the security of end-user devices.

The authors in [12], suggested AccessMiner(AM), a system that uses system-centric models to study software behavior and spot ma licious activity.

A system-centric model of how software behaves on a device is built by AM, which then uses this model to find anomalies that could indicate malicious behavior. The system constructs models of typical software behavior by employing ML algorithms and a combination of static and dynamic analysis methods to extract features from software samples.

Using a large dataset of both benign and malicious software sam ples, the authors assess AM's performance. The study demonstrates that AM outperforms conventional methods in terms of both effi ciency and accuracy when it comes to malware detection[12].

System-centric models, static and dynamic analysis, and machine learning are combined in the study's algorithm. The study relies on a substantial set of examples of both benign and malicious software. The malware detection system's accuracy, expressed as the propor tion of benign and malicious software samples correctly classified, is the performance metric used to evaluate the outcomes.

In [13], the authors presented a smart approach for detecting An droid malware in a large dataset. They utilized some of the most pop ular android datasets such as VirusTotal[18], Marvin[17], Drebin[21], and Malgenome[19][20]. The authors propose an ML-based approach that utilizes requested permissions by an android app for malware detection. The paper identified a list of sensitive permissions which are not supposed to be requested by any user applications but rather should be only used by system apps.

The same group extended their work, proposing a method for de tecting Android malware utilizing API calls[14]. The proposed ap proach involves creating a feature vector based on API calls and permissions, which are then used to train an ML classifier. The per formance of the proposed method was evaluated on a large dataset, and the results showed improved accuracy compared to existing ap proaches[13][14]. The authors conclude that the combination of API calls and permissions (check figures 1 and 2 for a list of sensitive APIs and permissions) can be used as a robust and effective feature set for detecting malware on Android devices. The performance was evaluated using several metrics, such as accuracy, precision, recall, and F1-score. The results show that the proposed approach outper forms other state-of-the-art methods[14], achieving an accuracy of 99.08%, a precision of 98.55%, a recall of 99.20%, and an F1-score of 98.87%.

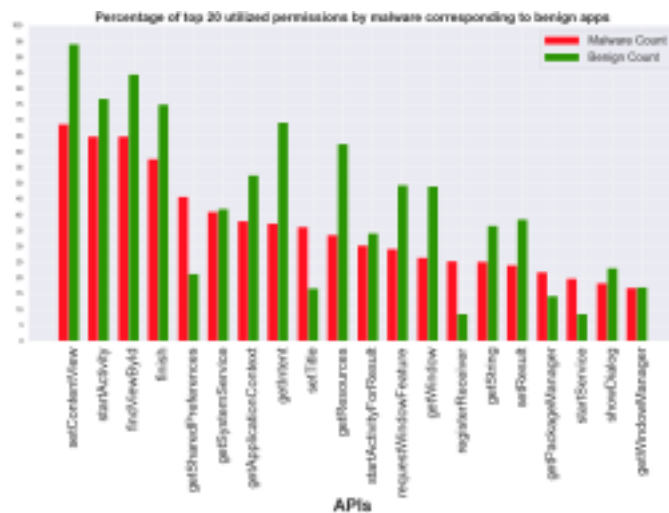

Fig. 1. List of sensitive APIs[14]

The static analysis involves extracting features from the Android Manifest and the Dalvik Bytecode, while the dynamic analysis in volves capturing system calls and network behavior. The dataset used for evaluation consists of over 10,000 Android applications, of which 5,000 are benign and 5,000 are malicious.

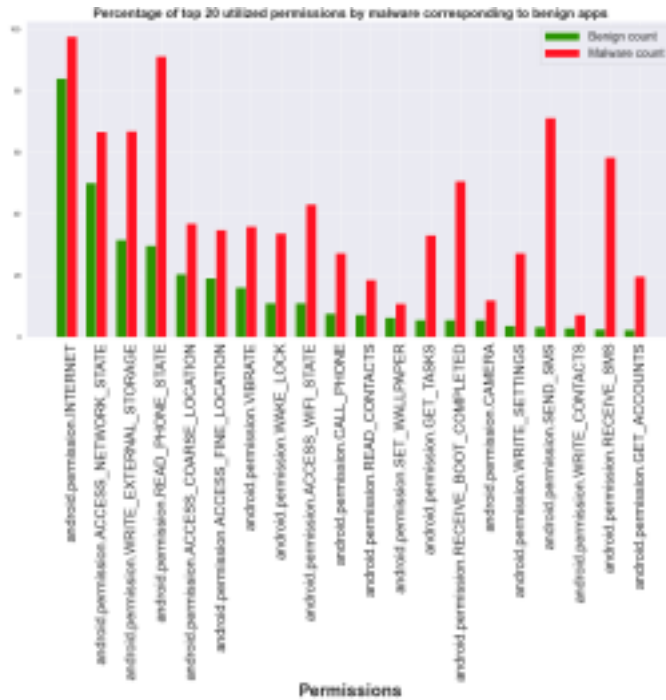

Fig. 2. List of sensitive permissions[13]

In [15], their focus was on a hybrid deep learning model for An droid malware detection. They used LSTM[30] and CNN algorithms [31] with two datasets: one from AndroZoo[22] and the other from VirusShare[23]. In terms of accuracy and F1-score, the experiments show that the hybrid deep learning model outperforms conventional ML algorithms[32], demonstrating the method's efficacy for Android malware detection.

In [16], a deep learning-based Android malware detection system is presented. The authors made use of a two-phase deep learning model: the prediction training phase and the testing phase. The model is trained on a large dataset of both benign and malicious applications. The deep learning model is used to predict whether an unidentified Android application is malicious, during the prediction phase. Over 10,000 legitimate and malicious Android applications were used in the authors' dataset. The data came from Google Play, third-party marketplaces, malware databases[33], and other sources. The accuracy, precision, recall, F1-score, and Area Under the ROC Curve (AUC) were used by the authors to measure the MAPAS's (Malware Analysis and Protection Using Artificial Intelligence Sys tem) performance. The findings demonstrate that the MAPAS sys tem was able to identify Android malware with a high degree of accuracy—more than 98%. In addition, the system demonstrated high precision, recall, F1-score, and AUC, all of which indicate its effectiveness in detecting Android malware.

## 3 METHODOLOGY

### 3.1 Overview of the Selection Criteria

A set of selection criteria was established to guide the selection of the studies to be included in this survey. Such selection criteria will provide a comprehensive review of the various ML approaches that are utilized for the detection of Android malware. In addition, these selected criteria will provide a comprehensive overview of the current state of the art and their relevance to Android malware detection.

Among the criteria for selection are:

Relevance: This includes research on malware detection on An droid platforms using ML algorithms.

Year of publication: It is important to keep up with the latest developments, so studies published in recent years (since 2015) were given priority.

Methodology: For Android malware detection, the studies in cluded in this survey must make use of ML algorithms.
Evaluation: Quantitative metrics like accuracy, time, and ro
bustness are used in the evaluations of the ML algorithms
included in this survey.

Data availability: The studies that are included in this survey
either have to make the evaluation data available to the general
pub lic or provide enough information to make it possible to
reproduce the results.

## 3.2 Selection of the Papers

A comprehensive search was carried out using multiple sources, in
cluding Google Scholar and online databases like IEEE Xplore,
Springer,

ACM Digital Library, and ScienceDirect, to locate relevant studies
for this review. A set of keywords related to Android malware de
tection and machine learning were used in the search.

The initial search yielded a plethora of results, which were
then filtered according to the selection criteria outlined in the
preceding section. In order to determine each study's relevance
and suitability for inclusion in this survey, the abstract and full text
was thoroughly examined during this process.

## 3.3 Data Collection and Analysis

The selected papers were thoroughly examined during the
process of data collection and analysis to obtain pertinent
information on Android malware detection using ML. To ensure
that this review's findings are consistent, complete, and current,
this information was collected in a structured manner.

Each paper contained the following information that we
extracted. The goal of the study was to find common themes,
trends, and gaps in the existing literature. An overview of the
current state of the art in Android malware detection using ML,
including the ad vantages and disadvantages of the methods that
are in use, is made
possible by the results of this analysis in this paper. Additionally,
the data gathered from the selected papers were utilized for contrasting and contrasting the various approaches as well as determining potential areas of future study. With the help of this analysis, a comprehensive understanding of the field's current state was provided, as well as the main obstacles and opportunities for future research.

## 4 RESULTS AND DISCUSSION

### 4.1 Overview of the Key Findings

This section presents the main findings of this literature review on Android malware detection with ML. A comprehensive analysis of the selected papers, which were chosen based on the established se lection criteria, serves as the foundation for the findings. Next is a summary of the most important findings from this review.

Android malware detection frequently makes use of ML algo rithms in the majority of the studies examined in this paper. Hence,

we can say that ML is the appropriate workhorse for Android Mal ware detection.

For Android malware detection, a variety of ML algorithms are utilized. Various machine learning algorithms, such as decision trees, artificial neural networks, support vector machines, and others, are in the reviewed studies. Yet, depending on the system's particular requirements and the nature of the data being analyzed, different ML algorithms will vary in their performances to carry out the malware detection task.

The Android malware detection system's performance is highly dependent on the selected dataset. The selection of the dataset is crucial to the system's performance and can significantly affect the outcomes. A variety of datasets, both real-world and synthetic, were used in the reviewed studies.

The reviewed studies have a wide range of evaluation metrics. A variety of evaluation metrics, such as accuracy, precision, recall, and the F1-score, were utilized in the reviewed studies. The varying of such evaluation metrics emphasizes the significance of selecting the appropriate evaluation metric for the system's particular require ments.

## 4.2 Summary of the Contributions

Based on our comprehensive literature review on Android malware detection using machine learning, the following are the main contri butions made by this review:

1. A systematic review of relevant sources: The relevant literature on Android malware detection using machine learning is system atically examined in this review. The papers were chosen using the established selection criteria, and thorough and systematic data collection and analysis were carried out.
2. An overview of how Android malware is detected using machine learning: The various machine learning algorithms and datasets used in Android malware detection are covered in this paper of the use of machine learning. Hence, researchers and practitioners in the field seeking to comprehend the current state of the art in this field may find this information helpful.
3. Analyzing the advantages and disadvantages of current meth ods: The current machine learning-based methods for Android

    malware detection are compared and contrasted in this review. The review sheds light on the difficulties and drawbacks of these approaches and reveals the areas that require additional investi gation.
4. Identifying future directions for research: Future directions for machine learning-based Android malware detection research are identified in this review. The review offers suggestions for enhanc ing the performance of existing methods and developing new, more efficient methods for this task.

By providing a comprehensive overview of the current state of the art, evaluating the strengths and weaknesses of existing approaches, and identifying future research directions, this review makes a sig nificant contribution to the field of Android malware detection using machine learning. The paper's findings can be used to guide the cre ation of Android malware detection systems that are more effective and to advance future research in this field.

### 4.3 Discussion of the Limitations

Although the current review provides a comprehensive overview of the existing literature on the application of machine learning to the detection of Android malware, it does have some drawbacks. The following are some significant limitations.

1. The literature covered: The current review looks at the litera ture that has been written up to a certain point, so it might not include the most recent work on this subject. As a result, it's possible that this review missed out on some significant research or developments in this area.

2. Dataset with bias: The quality and composition of the datasets used to determine the effectiveness of machine learning algo rithms for Android malware detection. Numerous studies have used datasets that may not accurately represent the distribution of malware in the real world or may be biased toward partic ular types of malware[33]. The generalizability of these studies' findings may be limited as a result.

3. Standard metrics for evaluation are missing: The absence of a standard evaluation metric presents a significant obstacle when assessing the effectiveness of machine learning algorithms for An droid malware detection. It is difficult to compare the results of different studies because different metrics have been used in each one.

4. Demand for extensive and varied datasets: To accurately capture the patterns and characteristics of malware, ML algorithms for Android malware detection require extensive and diverse datasets. However, obtaining such datasets is difficult, and numerous previ ous studies have utilized smaller or less diverse datasets, limiting the algorithms' accuracy[33].

5. Malware for Android is complex: It is challenging to develop ef ficient ML algorithms for detecting Android malware because it is highly dynamic, i.e. constantly changing. Algorithms that are capable of adapting to shifts in the malware landscape and ac curately detecting all types of malware are difficult to develop because of this complexity.

Even though there are some limitations, this review's findings are a good place to start more research on Android malware detection with machine learning. The limitations provide insight into how to improve the performance of existing algorithms and how to develop more efficient algorithms for this task. They also highlight the areas in which additional research is required.

## 4.4 Identification of Future Research Directions

The following are some possible directions for future machine learning based Android malware detection research based on the following review's findings:

1. Improvement of diverse and more accurate datasets: The absence of extensive and diverse datasets is one of the greatest obstacles in the development of efficient machine learning algorithms for Android malware detection. Future research should focus on cre ating more diverse and accurate datasets that accurately repre sent the distribution of malware in the real world to address this issue more accurately.
2. Utilization of deep learning methods: Convolutional neural net works (CNNs) and recurrent neural networks (RNNs) are two ex amples of deep learning methods that have demonstrated promis ing results in numerous applications, including speech and image recognition[31]. The focus of future research should be on advanc ing these two techniques and identifying applications where they outperform all the other peers.
3. The creation of adaptive and dynamic algorithms: It is very chal lenging to develop efficient ML algorithms for detecting Android malware because of its highly dynamic and constantly chang ing applications' environments. The development of dynamic and adaptable algorithms that can respond to shifts in the malware landscape ought to be the primary focus of future research.

4. Including security-related features: Code structure and API calls are two examples of features that have been used in numerous studies that are not specifically related to security. For Android malware detection, security-related features like permission re quests and system logs should be investigated, in more depth, in future research.

5. Evaluation of the algorithms in comparison: The absence of a standard evaluation metric presents a significant obstacle when assessing the effectiveness of machine learning algorithms for An droid malware detection. The development of a standard evalu ation metric and the comparative evaluation of algorithms that make use of this metric should be the primary focus of subsequent research.

6. Integration with current security measures: ML-based Android malware detection can be integrated with existing security sys tems to offer greater protection against malware. The effective ness of these algorithms and their integration with existing se curity systems should be investigated and evaluated in future subsequent research.

In summary, there is a lot of room for additional research in the field of Android malware detection using ML, which is rapidly evolv ing. This review's future research directions will help advance the field and enhance the effectiveness of Android malware detection al gorithms and serve as a useful starting point for additional research.

## 5 CONCLUSION

Malware for Android has become a serious threat to the Android platform's and its users' security, in recent years. Android malware detection has become a vital area of research due to the rapid growth of mobile devices and the ease with which malicious software can be distributed by intruders. ML-based solutions have been proposed and implemented to address this critical issue. In this paper, we conducted a comprehensive literature review on the use of ML to smartly detect Android malware. Our objective was to provide a comprehensive understanding of the current state of the art in this field, highlight the limitations and shed some light on future research directions, and highlight the most important findings and contribu tions of the most recent related research in the field.

Through our comprehensive review of the relevant literature, we found out that ML has been extensively used for Android malware detection and has been demonstrated to be effective in detecting mal ware in numerous instances. Decision trees, random forests, support vector machines, artificial neural networks, and deep learning-based strategies are among the ML algorithms that have been utilized for this purpose. System calls, API calls, and permissions are among the feature sets that have been used as input for training these al gorithms.

Additionally, our literature review revealed that much more re search is required to address some of the current approaches' draw backs. For instance, the generalizability of many of the existing meth ods to new and evolving malware is poorly understood because they are only tested on a small number of malware types. Additionally, more in-depth evaluations of these approaches are required, with an increased focus on the trade-off between efficiency and accuracy.

In conclusion, the current state of the art in Android malware de tection using machine learning is comprehensively reviewed in this paper. This survey's significant findings and contributions offer re searchers and practitioners in the field valuable insights. This study's limitations and future research directions serve as a road map for fu ture research in this field. We believe that this paper will be a very useful reference for those who are interested in this field. Such belief is based on the ongoing development of effective and efficient ML based solutions to detect and prevent Android malware, which is a crucial area of research with practical significance.